\documentclass{aa}
\usepackage{graphicx}

\begin{document}

\title{Point X-ray sources in the SNR G\,315.4$-$2.30 (MSH\,14$-$6{\it 3},
RCW\,86)}

\author{V.V.\,Gvaramadze
\inst{1,2,3}\thanks{{\it Address for
correspondence}: Krasin str. 19, ap. 81, Moscow, 123056, Russia
(vgvaram@sai.msu.ru)}
\and
A.A.\,Vikhlinin
\inst{4}
}
\institute{Sternberg Astronomical Institute, Moscow State
University, Universitetskij Pr.~13, Moscow, 119992,
Russia
\and
E.K.Kharadze Abastumani Astrophysical Observatory, Georgian Academy of
Sciences, A.Kazbegi ave.~2-a, Tbilisi, 380060, Georgia
\and
Abdus Salam International Centre for
Theoretical Physics, Strada Costiera~11, P.O.\ Box 586, 34100 Trieste,
Italy
\and
Space Research Institute, Russian Academy of Sciences,
Profsoyuznaya~84/32, Moscow, 117997, Russia
}
\offprints{V.V.\,Gvaramadze}

\date{Received 1 August 2002 / Accepted 28 October 2002}

\titlerunning{Point X-ray sources in the SNR G\,315.4$-$2.30}
\authorrunning{Gvaramadze \& Vikhlinin}

\abstract{We report the results of a search for a point X-ray
source (stellar remnant) in the southwest protrusion of the
supernova remnant G\,315.4$-$2.30 (MSH\,14$-$6{\it 3}, RCW\,86)
using the archival data of the {\it Chandra X-ray Observatory}.
The search was motivated by a hypothesis that G\,315.4$-$2.30 is
the result of an off-centered cavity supernova explosion of a
moving massive star, which ended its evolution just near the edge
of the main-sequence wind-driven bubble. This hypothesis implies
that the southwest protrusion in G\,315.4$-$2.30 is the remainder
of a pre-existing bow shock-like structure created by the
interaction of the supernova progenitor's wind with the
interstellar medium and that the actual location of the supernova
blast center is near the center of this hemispherical structure.
We have discovered two point X-ray sources in the ``proper" place.
One of the sources has an optical counterpart with the
photographic magnitude $13.38\pm0.40$, while the spectrum of the
source can be fitted with an optically thin plasma model. We
interpret this source as a foreground active star of late spectral
type. The second source has no optical counterpart to a limiting
magnitude $\sim 21$. The spectrum of this source can be fitted
almost equally well with several simple models (power law: photon
index $=1.87$; two-temperature blackbody: $kT_1 =0.11$ keV, $R_1
=2.34 $ km and $kT_2 =0.71$ keV, $R_2 =0.06$ km; blackbody plus
power law: $kT =0.07$ keV, photon index $=2.3$). We interpret this
source as a candidate stellar remnant (neutron star), while the
photon index and non-thermal luminosity of the source (almost the
same as those of the Vela pulsar and the recently discovered
pulsar PSR J\,0205$+$6449 in the supernova remnant 3C\,58) suggest
that it can be a young ``ordinary" pulsar.

\keywords{Stars: neutron --
    ISM: bubbles --
    ISM: individual objects: G\,315.4$-$2.30 (MSH\,14$-$6{\it 3}, RCW\,86) --
    ISM: supernova remnants --
    X-ray: stars}
}

\maketitle

\section{Introduction}
%

\object{G\,315.4$-$2.30} (\object{MSH\,14$-$6{\it 3}},
\object{RCW\,86}) is a bright (in both radio and X-ray) shell-like
supernova remnant (SNR) with a peculiar protrusion to the
southwest (e.g. Dickel et al. \cite{dic01}, Vink et al.
\cite{vin97}). This protrusion encompasses a bright hemispherical
optical nebula (Rodger et al. \cite{rod60}, Smith \cite{smi97};
see also Fig.~\ref{arc}). The characteristic angular size of the
SNR is about $40^{'}$, that at a distance to the remnant of 2.8
kpc (Rosado et al. \cite{ros96}) corresponds to $\simeq 32$ pc.
The radius of the optical nebula is $\simeq 2^{'}$ (or $\simeq
1.6$ pc). A collection of observational data points to the young
age (few thousand years) of the SNR (see e.g. Dickel et al.
\cite{dic01}).

Vink et al. (\cite{vin97}) put forward the idea that the SNR
G\,315.4$-$2.30 is the result of a supernova (SN) explosion inside
a pre-existing wind-driven cavity (see also Dickel et al.
\cite{dic01}, Vink et al. \cite{vin02}) and noted that the
elongated shape of the SNR resembles that of a wind-driven cavity
created by a {\it moving} massive star (see Weaver et al.
\cite{wea77} and Brighenti \& D'Ercole \cite{bri94} for details).
On the other hand, it is believed that the origin of the southwest
protrusion is due to the interaction of the SN blast wave with a
density enhancement in the ambient interstellar medium. However,
as was correctly noted by Dickel et al. (\cite{dic01}), the
density enhancement (e.g. a high-density cloud) should result in a
concave dent in the shell, not in a protrusion. Dickel et al.
(\cite{dic01}) also suggested that this protrusion ``is perhaps
the key to what is going on".

We agree with the idea that G\,315.4$-$2.30 is a diffuse remnant
of an off-centered cavity SN explosion and supplement this idea by
the suggestion that the massive SN progenitor star exploded near
the edge of the main-sequence bubble. This suggestion implies that
the southwest protrusion is the remains of a bow shock-like
structure created in the interstellar medium by the
post-main-sequence winds (see Sect.\,4 and Gvaramadze
\cite{gva02}; cf. Wang et al. \cite{wan93}) and that the SN
exploded near the center of this hemispherical structure. Given
the youth of the SNR and assuming a reasonable kick velocity for
the stellar remnant, one can expect that the stellar remnant
should still be within the protrusion. Motivated by these
arguments, Gvaramadze (\cite{gva02}) searched for a possible
compact X-ray source to the southwest of G\,315.4$-$2.30 using the
{\it ROSAT} archival data, but the moderate spatial resolution of
the {\it ROSAT} PSPC precluded detection of point sources against
the bright background emission of the SNR's shell.

In this paper we report the discovery of two point X-ray sources
near the center of the hemispherical optical nebula using the
archival Advanced CCD Imaging Spectrometer (ACIS) data of the {\it
Chandra X-Ray Observatory}. We interpret one of the sources as a
foreground active star of late spectral type, and the second one
as a candidate stellar remnant (neutron star).

Note that Vink et al. (\cite{vin00}) also reported the discovery
of a point X-ray source in the southwest half of G\,315.4$-$2.30,
at about $7^{'}$ from the geometrical center of the SNR. We recall
that Vink et al. (\cite{vin97}) suggested that the SN blast wave
in this SNR takes on the shape of the pre-existing elongated
cavity (created by the stellar wind of the moving SN progenitor
star). However, the initially spherical shape of the wind-driven
cavity could be significantly affected by the stellar motion only
if the massive star reaches the edge of the cavity and the stellar
wind starts to interact directly with the ambient interstellar
medium (e.g. Weaver et al. \cite{wea77}); this implies that the SN
explodes near the edge of the future (young) SNR (see Sect.\,4).
However, the source discovered by Vink et al. (\cite{vin00}) is
located too far from the edge of G\,315.4$-$2.30. Moreover, the
spectral characteristics of the source and the presence of a
possible optical counterpart suggest that it is an active star
rather than the stellar remnant associated with the SNR (Vink et
al. \cite{vin00}; see also Sect.\,3.1).

\section{Point X-ray sources}

\subsection{X-ray images}

Fig.~\ref{rcwfull} shows the {\it Chandra} (0.7$-$2 keV) image of
the highly structured southwest corner of the SNR G\,315.4$-$2.30.
Two point X-ray sources are clearly visible at the northwest of
the image at $\alpha _{2000} = 14^{\rm h} 40^{\rm m} 31{\fs}33,
\delta _{2000} = -62^{\circ} 38^{'} 22{\farcs}8$ and $\alpha
_{2000} = 14^{\rm h} 40^{\rm m} 31{\fs}05, \delta _{2000} =
-62^{\circ} 38^{'} 16{\farcs}9$. For both sources we measured a
FWHM of $1{\farcs}1$, which is consistent with the {\it Chandra}
point spread function at the off-axis angle $2\farcm5$.
Fig.~\ref{rcwcent} shows a close-up of the region around the
sources, labelled S (southern source) and N (northern source).
\begin{figure}
 \caption{{\it Chandra} image of the southwest region of
the SNR G\,315.4$-$2.30. Both X-ray sources are surrounded by a circle.
North is up and east is to the left.}
  \label{rcwfull}
\end{figure}
\begin{figure}
 \caption{The enlarged view of the region around the point X-ray
sources, labelled S (southern) and N (northern).}
  \label{rcwcent}
\end{figure}

\subsection{Optical images}

Fig.~\ref{arc} shows the image of the hemispherical optical nebula
to the southwest of the SNR G\,315.4$-$2.30 from the Digital Sky
Survey (DSS-2, red plates). Fig.~\ref{center} shows an enlarged
view of the region around the point X-ray sources. The source S is
in good positional agreement with a point-like object immersed in
a diffuse optical filament. This object is indicated in the HST
Guide Star Catalog as non-stellar, with a photographic magnitude
of $13.38\pm 0.40$. We believe, however, that it was misclassified
due to the effect of the background diffuse emission and in fact
it is a star.
\begin{figure}
  \caption{The DSS-2 image of the optical nebula southwest of
the SNR G\,315.4$-$2.30. The position of both point X-ray sources is
surrounded by a circle.}
  \label{arc}
\end{figure}
The source N has no optical counterpart to a limiting magnitude
$\sim 21$.
\begin{figure}
  \caption{The enlarged DSS-2 image of the region around the point
X-ray sources. The circles are centered at the position of X-ray
sources.}
  \label{center}
\end{figure}

\subsection{Spectral analysis}

The spectra of the point sources were extracted from circular
regions with radii of $1.5$ arcsec. The background spectrum was
taken from a circle with a radius of $4.5$ arcsec at 6.5 arcsec
northeast of the point sources. The PSF model shows that the
fraction of the source flux in the background region is
negligible. The spectral modeling was performed in the 0.5$-$10
keV energy range using the XSPEC spectra fitting package. The
estimates of the interstellar absorption, $N_{\rm H}$, towards
G\,315.4$-$2.30 are quite uncertain (cf. e.g. Vink et al.
\cite{vin97} with Vink et al. \cite{vin02}), so the spectra were
fitted with $N_{\rm H}$ as a free parameter. In one case, however,
we used the fixed value of $N_{\rm H}$ (see below).

The spectrum of the source S is shown in Fig.~\ref{sS};
\begin{figure*}
\resizebox{9cm}{!}{\includegraphics{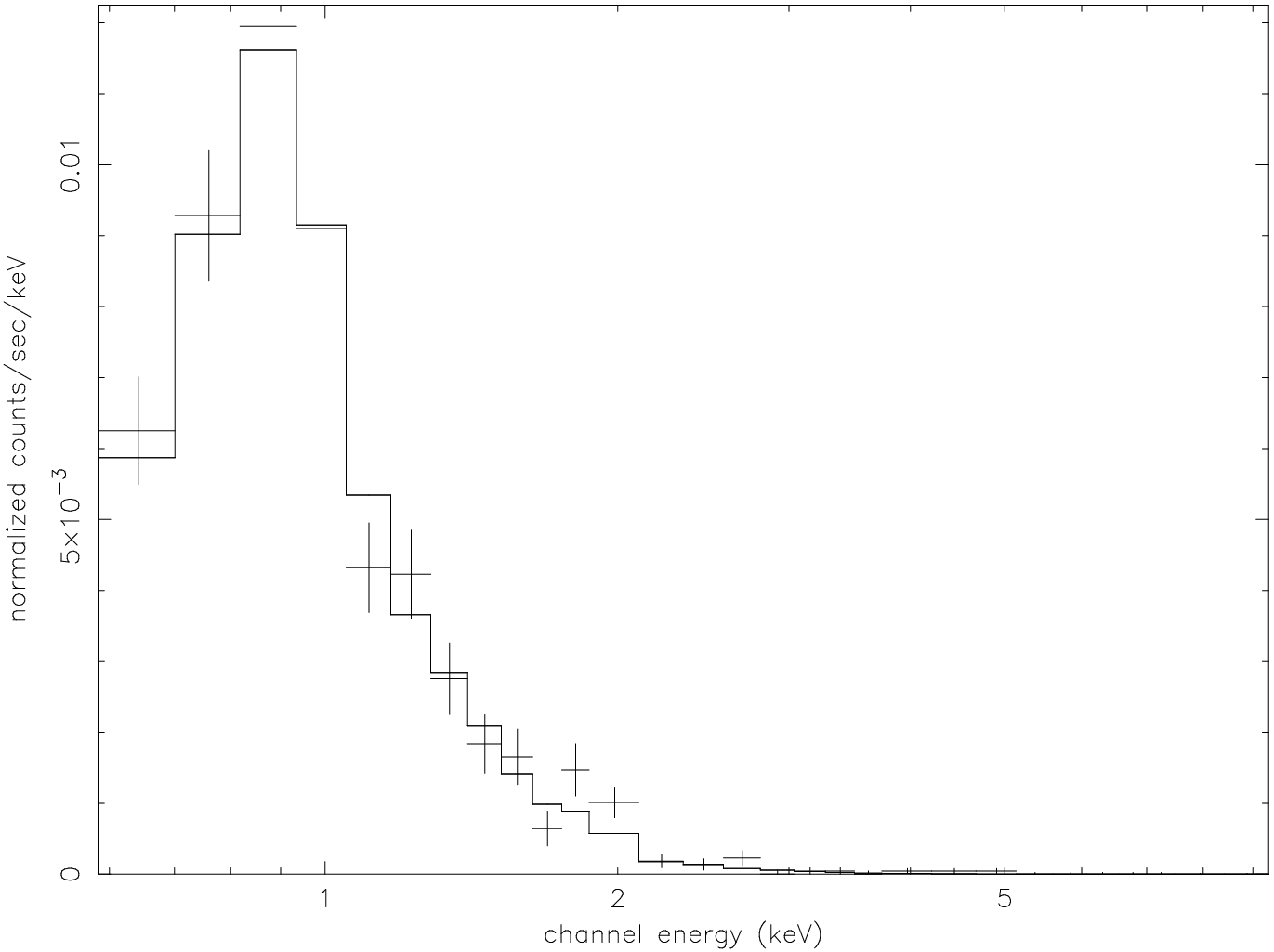}}
 \caption{The background-subtracted {\it Chandra} ACIS spectrum
from the source S. The solid line represents the best fit
optically thin plasma model.}
  \label{sS}
\end{figure*}
the solid line represents the best fit optically thin plasma
model, with a temperature $=0.69\pm 0.05$ keV, abundance $=0.07\pm
0.02$, $N_{\rm H} =(1.2\pm 0.3)\times 10^{21} \, {\rm cm}^{-2}$,
luminosity (in the 0.5-10 keV energy range) $\simeq 2.8\times
10^{31} \, {\rm erg} \, {\rm s}^{-1}$, and $\chi ^2$/d.o.f.
=21.6/25. The blackbody or power law models give unacceptable
fits.

\begin{figure*}
  \resizebox{9cm}{!}{\includegraphics{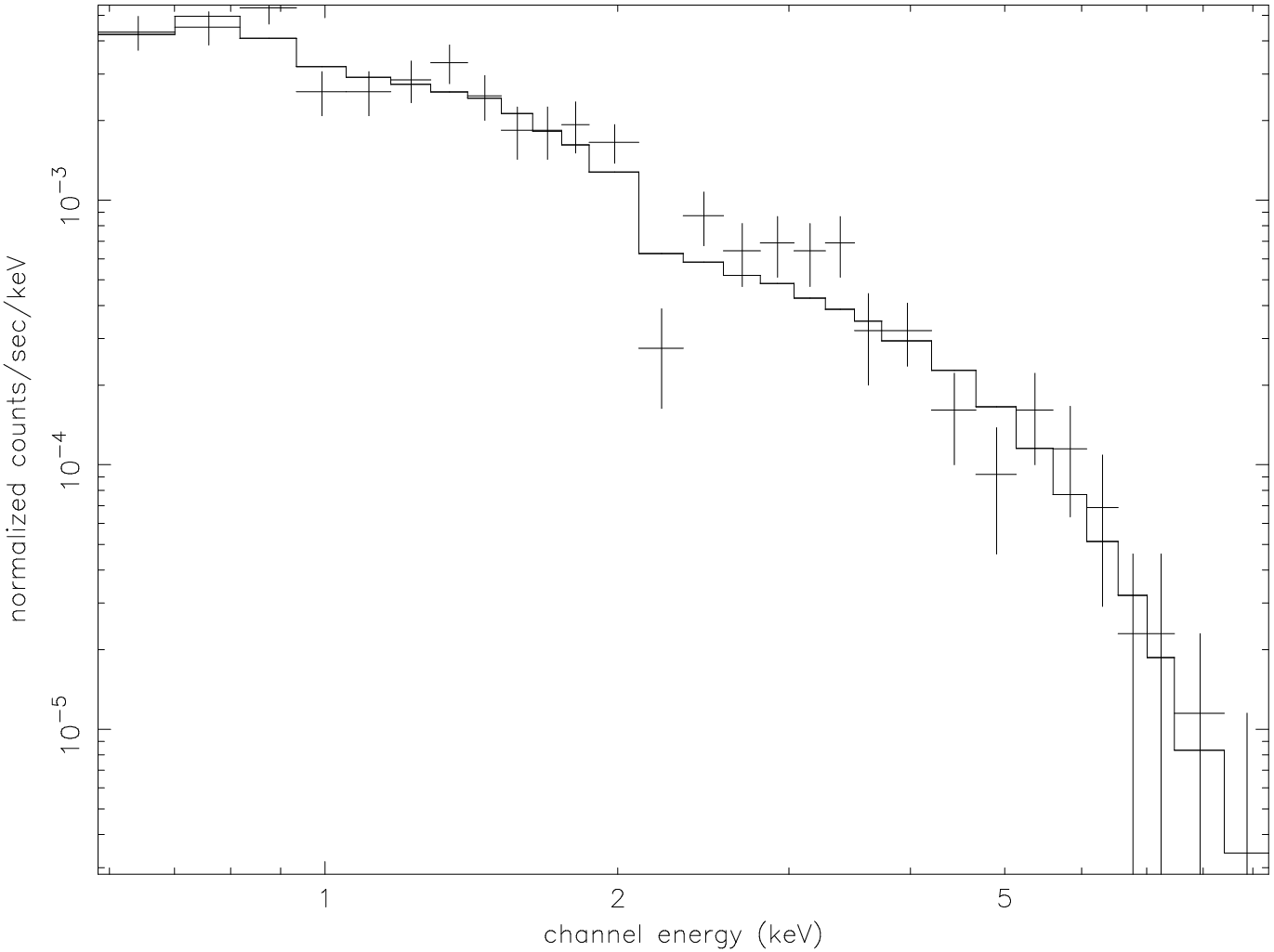}}
 \caption{The background-subtracted {\it Chandra} ACIS spectrum from
the source N. The solid line corresponds to the best fit blackbody plus
power law model.}
  \label{sN}
\end{figure*}

The spectrum of the source N (shown in Fig.~\ref{sN}) can be
fitted almost equally well with a power law (PL), two-temperature
blackbody (BB$+$BB), or blackbody plus power law (BB$+$PL) models.
A simple BB model gives unacceptable fits. The best fit spectrum
predicted by the BB$+$PL model is shown in Fig.~\ref{sN} by the
solid line.

We note that the best fit PL model requires $N_{\rm H} =0$ (we
consider this fact as an indication of the galactic origin of the source
N), although models with $N_{\rm H}$ up to $1.5\times 10^{21} \, {\rm
cm}^{-2}$ (presented in Table\,1) also give acceptable fits.

The results of a spectral analysis for the source N are summarized
in Table~\ref{table}.
\begin{table*}
  \caption[]{Spectral fits to the point X-ray source N.
\label{table}}
        \begin{flushleft}
                \begin{tabular}{llllll}
\hline\noalign{\smallskip}\hline\noalign{\smallskip} Model &
Photon index & Temperature, & $N_{\rm H},$ & Luminosity, & $\chi
^2$ /d.o.f.$^{\mathrm{a}}$\\
& & keV &  $ 10^{21} \, {\rm cm}^{-2}$ & $10^{32} \, {\rm erg} \,
{\rm s}^{-1}$ & \\
\noalign{\smallskip}\hline\noalign{\smallskip}\hline\noalign{\smallskip}
PL & $1.87\pm 0.09$ & & 1.5 (fixed) & 0.44 ($0.5-10\,{\rm keV}$) & 51.5/27\\
\noalign{\smallskip}\hline\noalign{\smallskip}
B$+$B & & $0.11 \pm 0.03$ & $4.3\pm 1.8$ & 1.32 (bolometric) & 47.3/25\\
 & & $0.71\pm 0.05$ & & 0.35 (bolometric) &\\
\noalign{\smallskip}\hline\noalign{\smallskip}
 BB$+$PL & &
$0.070\pm 0.003$ & $9.2\pm 3.4$ & 112 (bolometric) & 38.9/25\\
& $2.31\pm 0.30$ & & & 0.67 ($0.5-10 \, {\rm keV}$) &\\
\noalign{\smallskip}\hline
                \end{tabular}
        \end{flushleft}
\begin{list}{}{}
\item[$^{\mathrm{a}}$] Note that a single point near the Chandra mirror edge contributes
$\simeq 10$ to the
$\chi ^2$ in all cases.
\end{list}
\end{table*}

\section{Discussion}

\subsection{Source S}

The soft X-ray spectrum of the source S and the existence of the
candidate optical counterpart strongly suggest that this source is
an active star. Assuming that S is indeed a star and given the
observed X-ray to optical flux ratio, one can exclude the
possibility that this star is of the OB-type. Therefore it cannot
be a member of a group of OB stars located in the direction of
G\,315.4$-$2.30 nearly at the same distance as the SNR (Westerlund
\cite{wes69}). Most likely the source S is a foreground active
star of late spectral type. A follow-up study of this object could
clarify its nature.

Note that the point X-ray source discovered by Vink et al.
(\cite{vin00}) has much in common with the source S. The spectra
of both sources are soft and could be fitted with the optically
thin plasma model with depleted abundances (typical of late-type
stars; e.g. Giampapa et al. \cite{gia96}). Both sources have
candidate optical counterparts with almost the same photographic
magnitudes. In both cases the X-ray to optical flux ratios suggest
that the optical objects are late-type stars. Evidence for
long-term variability of the former source (Vink et al.
\cite{vin00}) also suggest that it is an active star.

\subsection{Source N}

The spectral characteristics of the source N coupled with the
absence of an optical counterpart allow us to consider it as a
candidate stellar remnant. We now discuss this possibility.

\subsubsection{Power law model}

The PL fit of the spectrum (with the photon index typical of
pulsars) suggest that the source N can be an active
(rotation-powered) neutron star. Assuming that N is a
pulsar\footnote{The $3.2 \, {\rm s}$ individual exposure time of
the ACIS instrument makes the search for short pulsations
impossible.} and using the empirical relationships between the
non-thermal X-ray and spin-down luminosities of pulsars (e.g.
Becker \& Tr\"umper \cite{bec97}, Possenti et al. \cite{pos02}),
one can estimate the latter, $\dot{E}\sim 10^{35} \, d_{2.8} ^2 \,
{\rm erg} \, {\rm s}^{-1}$, where $d_{2.8}$ is the distance to the
SNR in units of 2.8 kpc.

The inferred low spin-down luminosity can be considered as an
argument against the possibility that the source N is a young
fast-rotating pulsar with a standard magnetic field ($10^{11} -
10^{13}$ G). Therefore N can be either a young pulsar born with a
low surface magnetic field (e.g. Blandford et al. \cite{bla83},
Urpin et al. \cite{urp86}) and/or a large rotation period (e.g.
Spruit \& Phinney \cite{spr98}) or an aged pulsar with a decaying
magnetic field. The first possibility would deserve detailed
consideration if a candidate period of 12 ms found for the central
X-ray source in Cas\,A (Murray et al. \cite{mur02a}) is confirmed
by an independent timing analysis. The second one implies that N
is an old pulsar (of an arbitrary age) projected by chance on the
SNR G\,315.4$-$2.30.

Note, however, that the above-mentioned empirical relationships
are quite uncertain and should be used with caution. For example,
the 267 ms radio pulsar \object{PSR B\,1853+01} associated with
the SNR \object{W\,44} (Wolszczan et al. \cite{wol91}) has almost
the same spin-down luminosity ($\simeq 4\times 10^{35} {\rm erg}
\, {\rm s}^{-1})$ as that derived above for the source N, while
the characteristic age of the pulsar ($\simeq 2\times 10^4$ yr;
consistent with a range of ages derived for W\,44 using the
standard Sedov-Taylor model [see Smith et al. \cite{smi85}])
points to its relative youth. On the other hand, the pulsar PSR
B\,1853+01 could be much older if its high spin-down rate
($\dot{P} \simeq 0.2\times 10^{-12} \, {\rm s} \, {\rm s}^{-1}$)
is connected to the interaction between the pulsar's magnetosphere
and the dense circumstellar matter\footnote{The large dispersion
measure variations inherent to this pulsar (Jacoby \& Wolszczan
\cite{jac99}) could be considered as an indirect support of this
suggestion.} (cf. Gvaramadze \cite{gva01}). The age of W\,44 also
could be much greater if the origin of this mixed-morphology SNR
is due to the cavity SN explosion (cf. Gvaramadze \cite{gva02}).

The uncertainties inherent to the empirical relationships are more
obvious in the case of the pulsar \object{PSR J\,0205$+$6449},
recently discovered in the SNR \object{3C\,58}. The spin-down
luminosity of this young (characteristic age $\simeq 5400$ yr)
pulsar is $2.6\times 10^{37} \, {\rm erg} \, {\rm s}^{-1}$ (Murray
et al. \cite{mur02b}), while its (non-thermal; see Slane et al.
\cite{sla02}) X-ray luminosity is $\simeq 2.1\times 10^{32} \,
d_{2.6} ^2 \, {\rm erg} \, {\rm s}^{-1}$, i.e. about two orders of
magnitude less than that predicted by the empirical relationships.
The disparity between the predicted and observed non-thermal
luminosities is even higher in the case of another young pulsar --
the \object{Vela pulsar} (Pavlov et al. \cite{pav01}). Note that
the photon indices and non-thermal luminosities of these two
pulsars are almost the same as those predicted for the source N by
the PL (or BB$+$PL) model, while their characteristic ages are of
the same order of magnitude as the age of the SNR G\,315.4$-$2.30
inferred in Sect.\,4. Therefore one cannot exclude that the source
N is a young pulsar with ``ordinary" parameters. This should be
tested observationally.

\subsubsection{Two-temperature blackbody model}

In the two-temperature blackbody model, the origin of the soft
component can be attributed to the cooling surface of a neutron
star, while the hard component to the polar caps heated by the
backflow of relativistic particles (e.g. Wang et al.
\cite{wan98}).

The best-fit BB$+$BB model yields the temperature of the soft
component of $0.11$ keV and the effective radius of 2.3 km. The
inferred temperature is somewhat larger than that predicted by
standard cooling models (e.g. Page \cite{pag98}), while the radius
is significantly smaller than the radius of the neutron star. In
principle, one can expect that the use of realistic neutron star
atmosphere models (e.g. Pavlov et al. \cite{pav95}) would adjust
these parameters to acceptable values. But the interpretation of
the hard component is more problematic. The effective polar cap
area inferred from the best-fit BB$+$BB model ($\sim 10^8 \, {\rm
cm}^{-2}$) is too small and the observed temperature is too large
to be consistent with models for heating of polar caps (e.g. Wang
et al. \cite{wan98} and references therein). Therefore we consider
this model as unplausible and suggest that at least at high
energies the X-ray emission of the source N is non-thermal.

\subsubsection{Blackbody plus power law model}

The use of the BB$+$PL model assumes that the soft X-ray emission
comes from the entire surface of a cooling neutron star or from
some smaller hot areas, while the hard X-rays are due to the
non-thermal magnetospheric emission.

The best fit BB$+$PL model shows that the spectrum of the source N
is dominated by the non-thermal emission at energies $> 1$ keV,
and suggests the existence of a soft thermal component. However,
the unknown interstellar absorption and uncertainties due to the
time-dependent decrease in ACIS low-energy quantum efficiency make
the estimates of the flux and bolometric luminosity of the BB
component unconstrained. The best fit model with $N_{\rm H}$ as a
free parameter (see Table\,1) requires large interstellar
absorption and consequent large bolometric luminosity. Although
the large absorption, in principle, is consistent with values
derived by Vink et al. (\cite{vin02}) for the SNR G\,315.4$-$2.30,
the large inferred bolometric luminosity implies an uncomfortably
large effective radius of $\simeq 60$ km. The normalization of the
BB component, however, is plausible for lower values of $N_{\rm
H}$. A more detailed consideration of this problem will be
possible only with updated {\it Chandra} calibration at low
energies. Note that dense clumps of circumstellar matter (the
natural products of interaction between post-main-sequence winds
of the SN progenitor star; e.g. Gvaramadze \cite{gva01} and
references therein; see also Sect.\,4) could add a significant
contribution to the neutral hydrogen absorption towards the source
N, as well as possibly causing the enhanced absorption towards the
central compact source in \object{Cas\,A} (required by some model
fits of its spectrum; e.g. Murray et al. \cite{mur02a}).

Note also that at present we cannot definitively confirm or reject
the existence of a soft BB component. Therefore we consider the PL
model with a reasonable value of $N_{\rm H}$ as a good
approximation for the spectrum of the source N. The follow-up
multiwavelength observations of this source, including the search
for pulsed emission, long term variability, and radio, optical or
$\gamma$-ray counterparts, may provide crucial information for an
understanding of its nature and therefore are highly desirable.

\section{SNR G\,315.4$-$2.30}

We now briefly discuss a scenario for the origin of the SNR
G\,315.4$-$2.30 (the more detailed description will be presented
elsewhere). We note that our scenario has much in common with the
model proposed by Wang et al. (\cite{wan93}) to explain the origin
of large-scale structures around the \object{SN\,1987A}, and
suggest that G\,315.4$-$2.30 is an older version of the latter.

We believe that the SNR G\,315.4$-$2.30 is the result of a cavity
SN explosion of a moving massive star, which after the
main-sequence (MS) phase (lasting $\sim 10^7$ yr) has evolved
through the red supergiant (RSG) phase ($\sim 10^6$ yr), and then
experienced a short ($\sim 10^4$ yr) ``blue loop" (i.e. the
zero-age MS mass of the star was $15-20 M_{\odot}$). During the MS
phase the stellar wind (with the mechanical luminosity, $L$, of
$\sim 10^{35} \, {\rm erg} \, {\rm s}^{-1}$) blows a large-scale
bubble in the interstellar medium. The bubble eventually stalls at
radius $R_{\rm st} \sim 12 (L_{35}/n)^{1/2} \, {\rm pc}$ in time
$t_{\rm st} \sim 10^6 (L_{35}/n)^{1/2} \, {\rm yr}$, where $L_{35}
= L/10^{35} \, {\rm erg} \, {\rm s}^{-1}$ and $n$ is the number
density of the interstellar gas. The current radius of SNR
G\,315.4$-$2.30 is $\simeq 16$ pc. The motion of that star causes
it to cross the stalled bubble and start to interact directly with
the unperturbed interstellar gas. This happens at time $t_{\rm cr}
\simeq R_{\rm st}/v_{\star} = 2.4\times 10^6 (L_{35}/n)^{1/2}
v_{{\star},5} ^{-1} \, {\rm yr}$, where $v_{{\star},5}$ is the
stellar velocity in units of $5 \, {\rm km} \, {\rm s}^{-1}$. The
time is in agreement with the duration of the MS phase if
$v_{\star} \simeq 1 \, {\rm km} \, {\rm s}^{-1}$. In this case the
SN progenitor star enters in the RSG phase while it is near the
edge of the MS bubble. During the relatively short RSG phase, the
star loses most (two thirds) of its initial mass in the form of a
dense, slow wind. The interaction of the RSG wind with the
interstellar medium results in the origin of a bow shock-like
structure with a characteristic radius, $r$, determined by the
relationship: $\dot{M} _{\rm RSG} v_{\rm RSG} /4\pi r^2 \simeq
n(2kT + m_{\rm H} v_{\star} ^2)$, where $\dot{M} _{\rm RSG}$ and
$v_{\rm RSG}$ are, correspondingly, the mass-loss rate and wind
velocity during the RSG phase, $T$ is the temperature of the
ambient interstellar medium, $k$ is the Boltzmann constant, and
$m_{\rm H}$ is the mass of a hydrogen atom. For $n\simeq 1\, {\rm
cm}^{-3}$ (Smith \cite{smi97}, Ghavamian et al. \cite{gha01}),
$\dot{M} _{\rm RSG} = 10^{-5} \, M_{\odot} \, {\rm yr}^{-1}$,
$v_{\rm RSG} = 10 \, {\rm km} \, {\rm s}^{-1}$, $T=8000$ K, and
$v_{\star} = 1 \, {\rm km} \, {\rm s}^{-1}$, one has $r\simeq 1.5$
pc. This value is in a comfortable agreement with the radius of
the hemispherical optical nebula in the southwest of
G\,315.4$-$2.30 ($\simeq 1.6$ pc).

About $\sim 10^4$ yr before the SN explosion the progenitor star
becomes a blue supergiant whose fast wind sweeps the material of
the RSG wind. We speculate that at the moment of SN explosion the
blue supergiant wind was trapped in the southwest direction by the
dense material of the bow shock-like structure, while in the
opposite direction it was able to break out in the low-density MS
bubble, so that the SN explodes inside a ``hollow" hemispherical
structure open towards the MS bubble.

The SN blast wave expands almost freely across the low-density MS
bubble (for the explosion energy of $10^{51}$ erg and mass of the
SN ejecta of $\simeq 3.5 \, M_{\odot}$, the expansion velocity of
the blast wave is $v_{\rm bl} \simeq 5\,000 \, {\rm km} \, {\rm
s}^{-1}$) and reaches the northeast edge of the bubble after $\sim
4.7\times 10^3$ yr. The density jump at the edge of the bubble
results in the abrupt deceleration of the blast wave to a velocity
of $\sim (\beta n_{\rm b} /n)^{1/2} v_{\rm bl} \sim 600-800 \,
{\rm km} \, {\rm s}^{-1}$ (the values derived from studies of
Balmer-dominated filaments encircling the SNR G\,315.4$-$2.30;
Long \& Blair \cite{lon90} and Smith \cite{smi97}), where $\beta
\simeq 5$ (e.g. Sgro \cite{sgr75}). This deceleration implies a
density jump of a factor of $\simeq 200-300$ or a number density
of the MS bubble gas $n_{\rm b} \simeq 0.003-0.005 \, {\rm
cm}^{-3}$, i.e. a reasonable value given the mass lost during the
MS phase is $\sim 0.5 \, M_{\odot}$.

In the southwest direction, however, the expansion of the SN blast
wave is hampered by the dense hemispherical circumstellar shell.
The shocked remains of this shell are now seen as the bright
southwest protrusion (shown in Fig.~\ref{arc}). The existence of
radiative filaments in this corner of the SNR implies that the
blast wave slows down to the velocity of $\simeq 100 \, {\rm km}
\, {\rm s}^{-1}$, that in its turn implies the density contrast of
$\sim 15\,000$. For $n_{\rm b} =0.003 \, {\rm cm}^{-3}$, one has
the number density of the circumstellar material of $\sim 50 \,
{\rm cm}^{-3}$, i.e. in a good agreement with the density estimate
derived by Leibowitz \& Danziger (\cite{lei83}). On the other
hand, the recent discovery of Balmer-dominated filaments
protruding beyond the radiative arc (see Fig.\,3 of Smith
\cite{smi97} and Fig.\,4 of Dickel et al. \cite{dic01}) suggests
that the SN blast wave has partially overrun the clumpy
circumstellar shell (cf. Franco et al. \cite{fra91}) and now
propagates through the interstellar medium. We believe that this
effect is responsible for the complicated appearance of the
southwest corner of the SNR. For the radial extent of protrusions
of $\simeq 2$ pc, and assuming that their mean expansion velocity
is $\sim 700 \, {\rm km} \, {\rm s}^{-1}$, one has that the
blowouts occured $\sim 3\times 10^3$ yr ago.

\section{Conclusion}

We have analyzed the archival {\it Chandra X-Ray Observatory} data
on G\,315.4$-$2.30 to search for a stellar remnant in the
southwest corner of this SNR. The search was motivated by the
hypothesis that the SNR G\,315.4$-$2.30 is the result of an
off-centered cavity SN explosion of a moving massive star, which
ends its evolution just near the edge of the main-sequence
wind-driven bubble. This hypothesis implies that the southwest
protrusion in G\,315.4$-$2.30 is the shocked material of a
pre-existing circumstellar structure and that the actual location
of the SN blast center is near the center of this structure. We
have discovered two point X-ray sources in the ``proper" place.
One of the sources is interpreted as a foreground active star of
late spectral type, while the second one as a candidate neutron
star (perhaps a young ``ordinary" pulsar). The follow-up
observations of these sources will help us to understand their
nature and thereby to test the hypothesis of the origin of the SNR
G\,315.4$-$2.30.

\begin{acknowledgements}
We are grateful to J.R.Dickel (the referee) for useful suggestions and
comments.
\end{acknowledgements}

\end{document}